# Detecting and resolving spatial ambiguity in text using named entity extraction and self learning fuzzy logic techniques

-V. R. Kanagavalli[1], Dr. K. Raja[2]

[1]Research Scholar, Sathyabama University, Chennai, kanagavalli.teacher@gmail.com
[2]Principal, Narasu's Sarathy Institute of Technology, Salem, raja_koth@yahoo.co.in

**Abstract**

Information extraction identifies useful and relevant text in a document and converts unstructured text into a form that can be loaded into a database table. Named entity extraction is a main task in the process of information extraction and is a classification problem in which words are assigned to one or more semantic classes or to a default non-entity class. A word which can belong to one or more classes and which has a level of uncertainty in it can be best handled by a self learning Fuzzy Logic Technique. This paper proposes a method for detecting the presence of spatial uncertainty in the text and dealing with spatial ambiguity using named entity extraction techniques coupled with self learning fuzzy logic techniques.

## I. Introduction

In today's globalized and much connected world there is a lot of importance associated with the spatial awareness and reasoning. Spatial information is abundantly found in various texts. The main issue with handling spatial information is a degree of imprecision that is always associated with it. Natural language has a remarkable feature of describing spatial relation information without using numbers and they also offer a variety of lexical terms for describing spatial location of entities. The main issue in the use of natural language for spatial descriptions is that there is no unique way of doing it and the statement context often plays a significant role in the spatial reasoning.

In domains such as medical texts and in robotics imprecision has to be dealt with even for relationships that are well defined in a mathematical sense, such as distances. Modelling qualitative spatial relations depends on properties such as the asymmetry of some expressions, the non-bijective relation between language and spatial concepts, the interaction between distances and orientation etc. Fuzzy logic is naturally oriented to handle imprecision and is hence very much suitable for handling the spatial ambiguity or uncertainty. Combined with self learning techniques it is a natural choice for handling spatial uncertainty and spatial queries in natural language that are inherently vague in nature. This paper proposes the use of fuzzy logic to identify spatial ambiguity from text during the process of named entity extraction.

## II. Information Extraction

An information extraction system generally converts unstructured text into a form that can be loaded into a database table. Useful information such as the names of people, places, or organization mentioned in the text is extracted without a deep understanding of the text. While information retrieval deals with the problem of finding relevant documents in a collection, information extraction identifies useful and relevant text in a document.

Methods to find the names of organizations or people from unstructured text will have some errors that are then propagated in the database; this makes the task of consistently extracting accurate semantic information from all types of text difficult and cumbersome [23].

The authors in [3] categorize the text into various types, namely, instructional (text books, manuals, etc.,), imaginative (advertisements, simple texts, etc.,) and informative (e.g., literary papers, patent documents).

## III. Named Entity Extraction

The term "Named Entity", was coined for the Sixth Message Understanding Conference (MUC-6)[5]. Named Entity Recognition and Classification (NERC) was recognized as one of the most important sub-tasks of Information extraction and included the recognition of information units like names, including person, organization and location names, and numeric expressions including time, date, money and percent expressions from unstructured text, such as newspaper articles.

In [4] the authors present an exhaustive collection of the works done in the field of Named Entity Recognition. The impact of textual genre (journalistic, scientific, informal, etc.) and domain (gardening, sports, business, etc.) has been rather neglected in the NERC literature. Few studies are specifically devoted to diverse genres and domains.

The most studied types are three specializations of "proper names": names of "persons", "locations" and "organizations", collectively known as "enamex" since the MUC-6 competition. The type "location" can in turn be divided into multiple subtypes of "fine grained locations": city, state, country, etc. ([6], [7]).

The authors in [8] [9] relax the types to be extracted and hence it is called as "open domain" NERC. In this line of research, [16] defined a named entity hierarchy which includes many fine grained subcategories, such as museum, river or airport, and adds a wide range of categories, such as product and event, as well as substance, animal, religion or colour.

**IV. Existing Techniques of Named Entity Extraction**

The named extraction methods can be broadly classified into two categories namely rule based models and machine learning methods [1].

The advantage of Rule-based approaches is that it could effectively exploit human knowledge and can be tuned conveniently. On the other hand, machine learning approaches, such as *maximum entropy* or *support vector machine*, is more independent from languages and simple to implement. Rule-based approaches slightly outperform machine learning ones in MUC-7 tests [1].

The current dominant technique for addressing the NERC problem is supervised learning. SL techniques include Hidden Markov Models (HMM), Decision Trees, Maximum Entropy Models (ME), Support Vector Machines (SVM) and Conditional Random Fields (CRF). These are all variants of the SL approach that typically consist of a system that reads a large annotated corpus, memorizes lists of entities, and creates disambiguation rules based on discriminative features. The main technique for Semi Supervised Learning is called "bootstrapping" and involves a small degree of supervision, such as a set of seeds, for starting the learning process. The typical approach in unsupervised learning is clustering. For example, one can try to gather named entities from clustered groups based on the similarity of context. There are other unsupervised methods too. Basically, the techniques rely on lexical resources (e.g., WordNet), on lexical patterns and on statistics computed on a large unannotated corpus.

For some application, the constraint of exact match is unnecessarily stringent. For instance, in some bioinformatics work, the goal is to determine whether or not a particular sentence mentions a specific gene and its function. Exact NE boundaries are not required: all is needed is to determine if the sentence does refer to the gene [19].

The efficiency of the Named Entity Recognition technique is measured by a final score called Entity Detection and Recognition Value (EDR).

**V. Spatial Uncertainty**

Obtaining spatial awareness from raw input text has been a primary research area in the field of situation awareness. The main issues faced by researches in this area are modelling/ representation, even extraction and disambiguation, querying, reasoning and visualization [20]. Spatial descriptions found in free text are mostly based on human perceptions and perceptions are intrinsically imprecise.

The qualitative approach to the representation of spatial knowledge is heavily inspired by the way spatial information is expressed verbally. The spatial propositions might correspond to several different spatial relations depending on the context in which they are used. This leads to a degree of spatial uncertainty. One of the sources of uncertainty is the need to express a certain relationship independently of the context in which it occurs [21].

## VI. Fuzzy Logic

Fuzzy logic was first developed by Lotfi Zadeh [10] for solving decision making problems with "IF-THEN" rules, later, it was used to deal with uncertainly and imprecise data management. For several years fuzzy logic has been applied in different filed, such as automated control ,consumer products, [11] industrial systems, automotives, decision analysis, medicine, geology, controlling aircraft flight, chemical reactor and processes, and applications of artificial intelligence such as expert systems, pattern recognition and robotics. The generality of fuzzy logic is needed to deal with complex problems in the realms of search. Fuzzy logic provides a foundation for the development of new tools for dealing with natural languages and knowledge representation, such as precipitated natural language, theory of hierarchical definability, and unified theory of uncertainty [12]. Fuzzy logic is an easy way to reach definitive conclusions based on vague, and ambiguous and imprecise and noise, therefore, it makes sense to control the individual to take a quick decision [13], [14].

The semantics of natural languages and information analysis is best handled by the epistemic facet of Fuzzy Logic. In the epistemic Facet, natural language is viewed as a system for describing perceptions and an important branch of the same is possibility theory and computational theory of perceptions [15].

## VII. Named Entity Extraction and Fuzzy logic

The recognition of *genuine names* is basically a fuzzy decision problem to computers. There is no exact right or wrong answer for a string to be a name. The only problem is how likely it is. Fuzzy values represent strings' likelihood or properness to be a name. Names are composed of several characters. There are several ways to transform the member characters' fuzzy value to the string's fuzzy value. The authors in [2] propose a fuzzy logic method to identify isolated Arabic characters from a document. To simplify data collecting and training, unigram models are adopted in their work. Additionally, some supplementary information such as positional feature is exploited to support statistical models.

Several models are used to estimate the fuzzy value of a string from the statistic data based on characters. These models include Markov models, bi-gram models, unigram models, etc. Depending on the amounts of features of different types of named entities are varied, the type of the model to be used is decided.

Also, Fuzzy logic is used for Named Entity Extraction Look up operation. Second, candidate words can be "fuzzy-matched" against the reference list using some kind of threshold edit-distance [17] or Jaro-Winkler [18]. This allows capturing small lexical variations in words that are not necessarily derivational or inflectional. For instance, *Frederick* could match *Frederik* because the edit-distance between the two words is very small (suppression of just one character, the 'c'). Jaro-Winkler's metric was specifically designed to match proper names following the observation that the first letters tend to be correct while name ending often varies.

## VIII. Proposed Method

This paper proposes the idea of using neuro-fuzzy reinforcement learning for classifying the spatial descriptors and expressions from free text documents. The basic idea behind fuzzy reinforcement learning is to apply a fuzzy partitioning scheme to the continuous state space and to introduce linguistic interpretation. The method proposed is the use of Adaptive Heuristic Critic (AHC) that typically consists of an action (or control) module and of a critic (or evaluation) module. There are two representative neuro-fuzzy reinforcement learning modules: GARIC (Generalized Approximate Reasoning for Intelligent Control) and RNN-FLCS (Reinforcement Neural –Network-based Fuzzy Logic Control System) [22].

The Named Entity Extraction technique can be used to extract the spatial components and also spatial descriptors from the unstructured text. The uncertain spatial description, which has a level of ambiguity in it, are treated as input to the self learning fuzzy system. The system has the ability to disambiguate the spatial descriptions and assign a fuzzy membership value to the spatial locations identified without ambiguity.

The basic idea proposed is to use the fuzzy predictor to predict the nature of the spatial descriptions used in the text to disambiguate the spatial uncertainty.

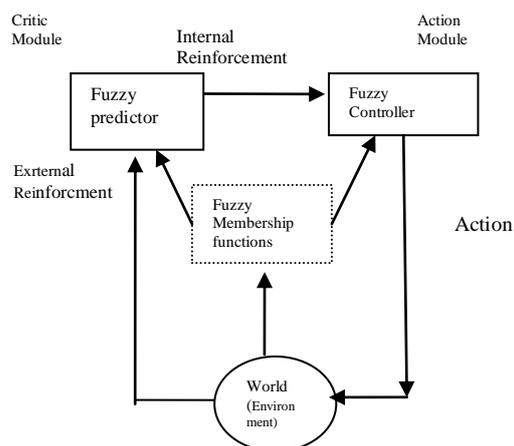

Fig 1. A Neuro-Fuzzy AHC Model

Fig.1 explains the overall working of a Neuro-Fuzzy AHC Model. Internal reinforcement is used to teach the system about the validity of the classification using the fuzzy membership functions. The action module consists of the fuzzy controller which classifies the ambiguous spatial descriptions.

## IX. Conclusion and Future Works

Though there are automated tools for extraction of spatial information from text there are still lot of errors and extraction ambiguities that creep in to the output. This paper proposes a self learning fuzzy system that would learn to identify the ambiguities and to resolve them. The so occurred spatial locations can then be further processed using fuzzy logic techniques for efficient and effective spatial querying in a large database setup.

## References

[1] Conrad Chen, Hsi-Jian Lee. A Three-Phase System for Chinese Named Entity Recognition

[2] Majida Ali Abed, Hamid Ali Abed AL-Asadi, Zainab Sabah Baha Al-Deen, Ahmad Naser Ismail. Fuzzy Logic approach to Recognition of Isolated Arabic Characters. International Journal of Computer Theory and Engineering, Vol. 2, No. 1 February, 2010. 1793-8201.

[3]. Ahmad, khurshid. Terminology and knowledge acquisition: A text based approach. In. Schmitz., Klaus-Dirk. (ed) 1993. Terminology and knowledge engineering, Proceedings. Third international congress on terminology and knowledge engineering 25-27.1993. 56-70.

[4] David Nadeau, Satoshi Sekine A survey of named entity recognition and classification. National Research Council Canada / New York University.

[5] Grishman, Ralph; Sundheim, B. 1996. Message Understanding Conference - 6: A Brief History. In *Proc. International Conference on Computational Linguistics*.

[6] Fleischman, Michael. 2001. Automated Subcategorization of Named Entities. In *Proc. Conference of the European Chapter of Association for Computational Linguistic*.

[7] Lee, Seungwoo; Geunbae Lee, G. 2005. Heuristic Methods for Reducing Errors of Geographic Named Entities Learned by Bootstrapping. In *Proc. International Joint Conference on Natural Language Processing*.

**[8]** Alfonseca, Enrique; Manandhar, S. 2002. An Unsupervised Method for General Named Entity Recognition and Automated Concept Discovery. In *Proc. International Conference on General WordNet*.

[9] Evans, Richard. 2003. A Framework for Named Entity Recognition in the Open Domain. In *Proc. Recent Advances in Natural Language Processing*.

[10] Zadeh ,Lotfi A. ,1998 , "Fuzzy logic Computer ", vol 22, issue 4.

[11] Zadeh ,Lotfi A. 2004 ,"Fuzzy logics systems: origin concepts ,and trends)

[12] Chen, G., Pham T., 2001 "Introduction to fuzzy sets, fuzzy logic, and fuzzy control systems ).

[13] Ruspini, E., 1992 "Introduction to Fuzzy set theory basic concepts And structures", IEEE.

[14] Brigette Krantz, A "Crisp" Introduction to Fuzzy Logic, Colorado University.

[15] Lotfi. A. Zadeh. Is there a need for fuzzy logic? Journal of information sciences.2008. 2751-2779.

[16] Sekine, Satoshi; Nobata, C. 2004. Definition, Dictionaries and Tagger for Extended Named Entity Hierarchy. In *Proc. Conference on Language Resources and Evaluation*.

[17] Tsuruoka, Yoshimasa; Tsujii, J. 2003. Boosting Precision and Recall of Dictionary-Based Protein Name Recognition. In *Proc. Conference of Association for Computational Linguistics Natural Language Processing in Biomedicine*.

[18] Cohen, William W.; Sarawagi, S. 2004. Exploiting Dictionaries in Named Entity Extraction: Combining Semi-Markov Extraction Processes and Data Integration Methods. In *Proc. Conference on Knowledge Discovery in Data*.

[19] Tzong-Han Tsai, Richard; Wu S.-H.; Chou, W.-C.; Lin, Y.-C.; He, D.; Hsiang, J.; Sung, T.-Y.; Hsu, W.-L. 2006. Various Criteria in the Evaluation of Biomedical Named Entity Recognition. *BMC Bioinformatics* 7:92, BioMed Central.

[20] Dimitri. V. Kalshinakov. SAT: Spatial awareness from textual input. International conference on Extending Database Technology, March 2006.

[21] Eliseo Climenteni et al. Qualitative representation of positional information. March 1997. Journal of Artificial Intelligence.

[22] J.S. R. Jang, C.T. Sun, E. Mizutani. Neuro-Fuzzy and Soft Computing. Prentice Hall International. Inc.,

[23] Manu Konchady. Text Mining Application Programming. Charles River Media. Boston. Massachusetts.